\documentclass[twocolumn,prd,aps,amsmath,amssymb,epsfig]{revtex4}

\usepackage{graphicx}% Include figure files
\usepackage{dcolumn}% Align table columns on decimal point
\usepackage{bm}% bold math

\begin{document}
\title{Polyakov Loops versus Hadronic States} \author{Francesco {\sc Sannino}}
\address{NORDITA, Blegdamsvej 17, DK-2100 Copenhagen \O, Denmark }
\date{April 2002}
%\maketitle

\begin{abstract}
The order parameter for the pure Yang-Mills phase transition is
the Polyakov loop which encodes the symmetries of the $Z_N$ center
of the $SU(N)$ gauge group. On the other side the physical degrees
of freedom of any asymptotically free gauge theory are hadronic
states. Using the Yang-Mills trace anomaly and the exact $Z_N$
symmetry we construct a model able to communicate to the hadrons
the information carried by the order parameter.
\end{abstract}
%]

\maketitle

\section{Introduction}
\label{uno}

Investigating the $SU(N)$ deconfinement phase transition is, in
general, a complex problem. At zero quark density importance
sampling lattice simulations are able to provide vital information
about the nature of the temperature driven phase transition for 2
and 3 colors Yang-Mills theories with and without matter fields
(see \cite{Boyd:1996bx,{Okamoto:1999hi}} for 3 colors). Different
approaches
\cite{Agasian:fn,Campbell:ak,Simonov:bc,Sollfrank:du,Carter:1998ti,Schaefer:2001cn,Drago:2001gd,Renk:2002md,Pisarski:2001pe,DP,KorthalsAltes:1999cp,Dumitru:2001xa,Wirstam:2001ka,Laine:1999hh,Sannino:2002re,Scavenius:2001pa,Scavenius:2002ru,DelDebbio:2002nb,Giacomo:2002qr}
are used in literature to tackle/study the features of this phase
transition.

At zero temperature $SU(N)$ Yang-Mills theory is asymptotically
free and the physical spectrum of the theory consists of a tower
of hadronic states referred as glueball and pseudo-scalar
glueballs. The theory develops a mass gap and the lightest
glueball has a mass of the order of few times the confining scale.
The classical theory is conformal while quantum corrections lead
to a non-vanishing trace of the energy momentum tensor.

At nonzero temperature the $Z_N$ center of $SU(N)$ is a relevant
global symmetry \cite{Svetitsky:1982gs} and it is possible to
construct a number of gauge invariant operators charged under
$Z_N$ among which the most notable one is the Polyakov loop:
\begin{eqnarray} {\ell}\left(x\right)=\frac{1}{N}{\rm Tr}[{\bf L}]\equiv\frac{1}{N}{\rm Tr}
\left[{\cal
P}\exp\left[i\,g\int_{0}^{1/T}A_{0}(x,\tau)d\tau\right]\right] .
\end{eqnarray} ${\cal P}$ denotes path ordering, $g$ is the $SU(N)$ coupling constant and $x$ is
the coordinate for the three spatial dimensions while $\tau$ is
the euclidean time. The $\ell$ field is real for $N=2$ while
otherwise complex.  This object is charged with respect to the
center $Z_N$ of the $SU(N)$ gauge group \cite{Svetitsky:1982gs}
under which it transforms as $\ell \rightarrow z \ell$ with $z\in
Z_N$. A relevant feature of the Polyakov loop is that its
expectation value vanishes in the low temperature regime while is
non zero in the high temperature phase. The Polyakov loop is a
suitable order parameter for the Yang-Mills temperature driven
phase transition \cite{Svetitsky:1982gs}.

This behavior recently led Pisarski \cite{Pisarski:2001pe} to
model the Yang-Mills phase transition as a mean field theory of
Polyakov loops. This model is often referred as the Polyakov Loop
Model (PLM). Using this model one can  infer that the $SU(2)$
phase transition is second order while a phase transition (as
function of the temperature) is a weak first order for $SU(3)$.
The predictions are in reasonable agreement with lattice results.
Moreover the PLM is used to model the Yang-Mills free energy.
Recently some interesting phenomenological PLM inspired models
aimed to understand RHIC physics were constructed
\cite{Scavenius:2001pa,{Scavenius:2002ru}}.

Here we will consider pure gluon dynamics. This allows us to have
a well defined framework where the $Z_N$ symmetry is exact. The
hadronic states are the glue balls fields ($H$) and their
effective theory at the tree level is constrained by the
Yang-Mills trace anomaly.

A real puzzle to me is how the information about the Yang-Mills
phase transition encoded, for example, in the $Z_N$ global
symmetry can be communicated to the hadronic states of the theory.
Here we propose a concrete model which can help resolving this
puzzle.

This model is constructed using trace anomaly and the $Z_N$
symmetry. We will demonstrate that the information carried by
$\ell$ is efficiently transferred to the glueballs.  More
generally the glueball field is a function of $\ell$:
\begin{eqnarray} H\equiv H[\ell] . \end{eqnarray} Our results
can be tested via first principle lattice simulations
\cite{Bacilieri:mj} and support the recent phenomenological
investigations \cite{Scavenius:2001pa,{Scavenius:2002ru}}.

 In section \ref{due} we present the model. In section \ref{tre}
we consider the two colors Yang-Mills theory while in
\ref{quattro} the three color theory is considered. We finally
conclude in section \ref{cinque}

\section{The Model}
\label{due} The hadronic states of the Yang-Mills theory are the
glueballs.  At zero temperature the Yang-Mills trace anomaly has
been used to constrain the potential of the lightest glueball
state $H$ \cite{Schechter:2001ts}:
\begin{eqnarray}
V[H]=\frac{H}{2} \ln \left[\frac{H}{\Lambda^4}\right] .
\end{eqnarray}
$\Lambda$ is chosen to be the confining scale of the theory and
$H$ is a mass dimension four field. This potential correctly
saturates the trace anomaly when $H$ is assumed to be proportional
to ${\rm Tr}\left[G_{\mu \nu}G^{\mu \nu}\right]$ and $G_{\mu \nu}$
is the standard Yang-Mills field strength. The potential nicely
encodes the properties of the Yang-Mills vacuum at zero
temperature and it has been used to deduce a number of
phenomenological results \cite{Schechter:2001ts}.

At high temperature Pisarski conjectured that the Yang-Mills
pressure can be written in terms of the field $\ell$. This free
energy must be invariant under $Z_N$ and it takes the general
form:
\begin{eqnarray}
V[\ell]= T^4{\cal F}[\ell] .
\end{eqnarray}
${\cal F}[\ell]$ is a polynomial in $\ell$ invariant under $Z_N$
and the coefficients depend on the temperature itself allowing for
a mean field description of the Yang-Mills phase transitions.

We now marry the two models by requiring both fields to be present
simultaneously at non zero temperature. The theory must reproduce
at zero and low temperatures the ordinary glueball Lagrangian
while the PLM at high temperatures. We propose the following
effective potential:
\begin{equation}
V\left[H,\ell \right]
=\frac{H}{2}\ln\left[\frac{H}{\Lambda^4}\right] +
V_{T}\left[H\right]+H{\cal P}{\left[ \ell \right]} + T^4 {\cal
V}\left[ \ell \right]   , \label{POTENTIAL}
\end{equation}
where ${\cal V}\left[ \ell \right]$ and ${\cal P}\left[ \ell
\right]$ are general (but real) polynomials in $\ell$ invariant
under $Z_N$ whose coefficients depend on the temperature. The
explicit dependence is not known and should be fit to lattice
data. Dimensional analysis and analyticity in $H$ when coupling it
with $\ell$ severely restricts the effective potential terms. We
stress that $H{\cal P}[\ell]$ is the most general interaction term
which can be constructed without spoiling the zero temperature
trace anomaly.

Further nonanalytic interaction terms can arise when considering
thermal and quantum corrections and are partially contained in
$V_{T}[H]$ which schematically represents the temperature of a gas
of glueballs. In the following we will not investigate in detail
such a term.  Our theory cannot be the full story since we
neglected (as customary) all of the tower of glueballs and
pseudo-scalar glueballs as well as the infinite series of
dimensionless gauge invariant operators with different charges
with respect to $Z_N$. Nevertheless the potential is sufficiently
general to hope to capture the essential features of the
Yang-Mills phase transition.

When the temperature $T$ is much less than the confining scale
$\Lambda$ the last term in Eq.~(\ref{POTENTIAL}) can be safely
neglected. Since the glueballs are relatively heavy compared to
the $\Lambda$ scale their temperature contribution $V_T[H]$ can
also be disregarded. At low temperatures the theory reduces to the
standard glueball potential augmented by the third term which does
not affect the trace anomaly.

At very high temperature (compared to $\Lambda$) the last term
dominates ($H$ itself is very small) recovering the picture in
which $\ell$ dominates the free energy. In this regime we have
${\cal F}[\ell]={\cal V}[\ell]$.

 We can, in principle, compute all of the relevant thermodynamical
quantities in our approach, i.e. entropy, pressure etc.

A relevant object is the trace of the energy-momentum tensor
$\Theta^{\mu}_{\mu}$. At zero temperature and when the potential
is a general function of a set of bosonic fields $\{\Phi_n \}$
with mass-dimensions $d_n$ one can construct the associated trace
of the energy-momentum tensor  via:
\begin{equation}
\Theta^{\mu}_{\mu}=4V[\Phi_n]-\sum_n\frac{\delta V[\Phi_n]}{\delta
\Phi_n}\Phi_n\, d_n .  \label{Ttheta}
\end{equation}
At finite temperature we still define our temperature dependent
energy-momentum tensor as in Eq.~(\ref{Ttheta}). Here $H$
possesses engineering mass dimensions $4$ while  $\ell$  is
dimensionless yielding the following temperature dependent stress
energy tensor:
\begin{equation}
\Theta^{\mu}_{\mu}(T)= -2H +4T^4{\cal V}[\ell] +
4\left[1-H\frac{\delta}{\delta H}\right]V_T [H] .
\end{equation}
$\Theta^{\mu}_{\mu}$ is normalized such that $ \langle 0|
\Theta^{\mu}_{\mu} |0\rangle = \epsilon - 3p $ with $\epsilon$ the
vacuum energy density and $p$ the pressure. At zero temperature
only the first term survives yielding magnetic type condensation
typical of a confining phase while at extremely high temperature
the second term dominates displaying an energy density and
pressure of the deconfined phase.

The theory containing just $\ell$  can be obtained integrating out
$H$ via the equation of motion:
\begin{eqnarray}
\frac{\delta V[H,\ell]}{\delta H}= 0 .   \label{OUT}
\end{eqnarray}
{}Formally this is justifiable if the glueballs degrees of freedom
are very heavy. {}For simplicity we neglect the contribution of
$V_T[H]$ as well as the mean-field theory corrections for $\ell$.
However in the future a more careful treatment which also includes
the kinetic terms should be considered. Within these
approximations the equation of motion yields:
\begin{eqnarray}
H[\ell]=\frac{\Lambda^4}{e}\exp\left[ -2 {\cal P}[\ell] \right] .
\label{Hofl}
\end{eqnarray}
The previous expression shows the intimate relation between $\ell$
 and the physical states of strongly interacting theories.

After substituting Eq.~(\ref{Hofl}) back into the potential
(\ref{POTENTIAL}) and having neglected $V_T[H]$ we have:
\begin{eqnarray}
V[\ell]=T^4 {\cal V}[\ell] - \frac{\Lambda^4}{2e}\exp \left[-2
{\cal P}[\ell] \right] .
\end{eqnarray}
This formula shows that for large temperatures the only relevant
energy scale is $T$ and we recover the PLM model. However at low
temperatures the scale $\Lambda$ allows for new terms in the
Lagrangian. Besides the $T^4$ and the $\Lambda^4$ terms we also
expect terms with coefficients of the type $T\Lambda^3$ and
$T^2\Lambda^2$ and $T^3 \Lambda$. However in our simple model
these terms do not seem to emerge.

Expanding the exponential we have:
\begin{eqnarray}
V[\ell]=T^4 {\cal V}[\ell] +\frac{\Lambda^4}{e}{\cal P}[\ell] -
\frac{\Lambda^4}{2 e} + \cdots .
\end{eqnarray}
Since ${\cal V}[\ell]$ and ${\cal P}[\ell]$ are real polynomials
in $\ell$ invariant under $Z_N$ we immediately recover a general
potential  in $\ell$.

\section{The two Color Theory}
\label{tre}
 To illustrate how our formalism
works we first consider in detail the case $N=2$ and neglect for
simplicity the term $V_T[H]$. This theory has been extensively
studied via lattice simulations \cite{Damgaard,{Hands:2001jn}} and
it constitutes the natural playground to test our model.  Here
$\ell$ is a real field and the $Z_2$ invariant ${\cal V}[\ell]$
and ${\cal P}[\ell]$ are taken to be:
\begin{eqnarray}
   {\cal V}\left[ \ell \right] &=& a_1 \ell ^2 + a_2 \ell^4  + {\cal
   O}(\ell^6) , \nonumber \\
     {\cal P}\left[ \ell \right] &=& b_1 \ell ^2 + {\cal
   O}(\ell^4) ,
\end{eqnarray}
with $a_1,~a_2$ and $b_1$ unknown temperature dependent functions
which should be derived directly from the underlying theory.
Lattice simulations can, in principle, fix all of the
coefficients. In order for us to investigate in some more detail
the features of our potential and inspired by the PLM model
mean-field type of approximation we first assume $a_2$ and $b_1$
to be positive and temperature independent constants while we
model $a_1=\alpha (T_{\ast}-T)/T$,  with $T_{\ast}$ a constant and
$\alpha$ another positive constant. We will soon see that due to
the interplay between the hadronic states and $\ell$,  $T_{\ast}$
need not to be the critical Yang-Mills temperature while $a_1$
displays the typical behavior of the mass square term related to a
second order type of phase transition.

The extrema are obtained by differentiating the potential with
respect to $H$ and $\ell$:
\begin{eqnarray}
\frac{\partial V}{\partial H}&=&
\frac{\ln}{2}\left[\frac{eH}{\Lambda^4}\right]+{\cal
P}\left[\ell\right]
=\frac{\ln}{2}\left[\frac{eH}{\Lambda^4}\right]+b_1\ell^2=0 , \label{motiona} \\
\frac{\partial V}{\partial \ell}&=&2\ell T^4\left(a_1 +
\frac{H}{T^4} b_1 + 2a_2 \ell^2 \right) =0  ,  \label{motionb}
\end{eqnarray}

\subsection{Small and Intermediate Temperatures}
At small temperatures the second term in Eq.~(\ref{motionb})
dominates and the only solution is $\ell=0$. A vanishing $\ell$
leads to a null ${\cal P}[\ell]$ yielding  the expected minimum
for $H$:
\begin{equation}
\langle H\rangle = \frac{\Lambda^4}{e} .  \label{Hvacuum}
\end{equation}
Here $\ell$ and $H$ decouple.

We now study the solution near the critical temperature for the
deconfinement transition. {}For all the temperatures for which
\begin{eqnarray}
T^4 a_1 + {H} b_1 =T^3 \alpha (T_{\ast}-T) + H b_1 > 0 ,
\end{eqnarray}
the solution for $\ell$ is still $\ell=0$ yielding
Eq.~(\ref{Hvacuum}). The critical temperature is reached for
\begin{eqnarray}
T_c = T_{\ast} + \frac{b_1}{e\alpha}\frac{\Lambda^4}{T_c^3} .
\label{criticalTc}
\end{eqnarray}
The critical temperature can be determined via lattice
simulations. We see that within our framework the latter is
related to the glueball (gluon-condensate) coupling to two
Polyakov loops and it would be relevant to measure it on the
lattice.  At $T=T_c$, $\ell=0$ and $H=\Lambda^4 /e$.

Let us now consider the case $T=T_c + \Delta T$ with
\begin{eqnarray}\frac{\Delta T}{T_c}  \ll 1 .\end{eqnarray}
Expanding $\langle \ell \rangle ^2$ at the leading order in
$\Delta T/T_c$ yields:
\begin{eqnarray}
\langle \ell \rangle^2 =\frac{\alpha}{2a_2}
\frac{1+3\frac{b_1}{e\alpha} \frac{\Lambda^4}{T_c^4}}{1 -
\frac{b_1^2}{ e a_2} \frac{\Lambda^4}{T_c^4}} \, \frac{\Delta
T}{T_c} .
\end{eqnarray}
We used Eq.~(\ref{criticalTc}) and Eq.~(\ref{motiona}) which
relates the temperature dependence of $H$ to the one of $\ell$. At
high temperatures (see next subsection) $\langle \ell \rangle $
can be normalized to one by imposing $\alpha/2a_2 =1$ and the
previous expression reads:
 \begin{eqnarray}
\langle \ell \rangle^2 = \frac{1+3\frac{b_1}{e\alpha}
\frac{\Lambda^4}{T_c^4}}{1 - \frac{2b_1^2}{e \alpha}
\frac{\Lambda^4}{T_c^4}} \, \frac{\Delta T}{T_c} \equiv\frac{4
T_c-3T_{\ast}}{(1-2b_1)T_c + 2b_1 T_{\ast}}\,\frac{\Delta T}{T_c}.
\end{eqnarray}
{}For a given critical temperature consistency requires $b_1$ and
$T_{\ast}$  to be such that:
\begin{eqnarray} \frac{4 T_c-3T_{\ast}}{(1-2b_1)T_c + 2b_1 T_{\ast}}
\geq 0 .
\end{eqnarray}  The temperature dependence, in this regime, of the gluon
condensate is:
\begin{eqnarray}
\langle H \rangle =\frac{\Lambda^4}{e}\, \exp\left[-2b_1 \langle
\ell \rangle ^2 \right]  .
\end{eqnarray}
We find the mean field exponent for $\ell$, i.e. $\ell^2$
increases linearly with the temperature near the phase transition
\footnote{Corrections to the mean field are large and must be
taken into account.}. Interestingly the gluon-condensates drops
exponentially. The drop in the gluon-condensate is triggered by
the rise of $\ell$ and it happens in our simple model exactly at
the deconfining critical temperature. Although the drop might be
sharp it is continuous in temperature and this is related to the
fact that the phase transition is second order. Our findings
strongly support the common picture according to which the drop of
the gluon condensate signals, in absence of quarks, deconfinement.

\subsection{High Temperature}
\begin{figure}[h]
\begin{center}
\includegraphics[width=7truecm]{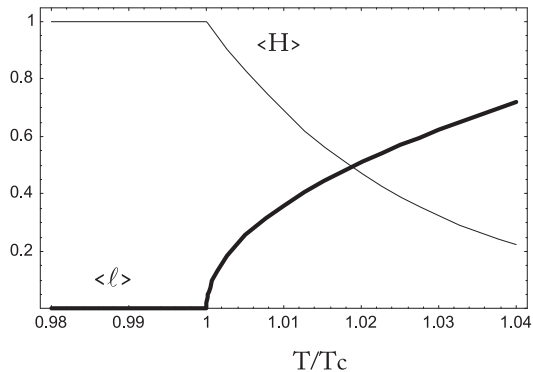}
\end{center}
\caption{The thin line is the gluon condensate $\langle H \rangle$
normalized to $\Lambda^4/e$ as function of the temperature. The
thick line represents the normalized to unity $\langle \ell
\rangle $ as function of the temperature. We have chosen for
illustration $\alpha=1$, $b_1=1.45$ and $T_c\simeq 1.16 \Lambda$.}
\label{Figura1}
\end{figure}

 At very high temperatures the second
term in Eq.~(\ref{motionb}) can be neglected and the minimum for
$\ell$ is:
\begin{equation}
\langle \ell \rangle =\sqrt{\frac{\alpha}{2a_2}}. \label{HighT}
\end{equation}
{}For $H$ we have now:
\begin{eqnarray}
\langle H \rangle = \frac{\Lambda^4}{e} \exp \left[
-2b_1\frac{\alpha }{2 a_2} \right] = \frac{\Lambda^4}{e} \exp
\left[ - 2b_1 \right]  .
\end{eqnarray}
In the last step we normalized $\langle \ell \rangle$ to unity at
high temperature. In order for the previous solutions to be valid
we need to operate in the following temperature regime:
\begin{eqnarray}
T \gg \sqrt[4]{\frac{b_1}{\alpha} {\langle H \rangle}}\approx T_c.
\end{eqnarray}
  We find that at sufficiently high temperature $\langle H \rangle$ is exponentially
suppressed and the suppression rate is determined solely by the
glueball -- $\ell^2$ mixing term encoded in ${\cal P}[\ell]$. The
coefficient $b_1$ should be large (or increase with the
temperature) since we expect a vanishing gluon-condensate at
asymptotically high temperatures. Clearly it is crucial to
determine all of these coefficients via first principle lattice
simulations. The qualitative picture which emerges in our analysis
is summarized in Fig.~\ref{Figura1}.

\section{The three color theory}
\label{quattro} $Z_3$ is the global symmetry group for the three
color case while $\ell$ is a complex field. The functions ${\cal
V}[\ell]$ and ${\cal P}[\ell]$ are:
\begin{eqnarray}
     {\cal V}\left[ \ell \right] &=& a_1 |\ell| ^2 +
     a_2 |\ell|^4  -a_3 (\ell^3 + {\ell^{\ast}}^3)+{\cal
   O}(\ell^5)  ,\nonumber \\
     {\cal P}\left[ \ell \right] &=& b_1 |\ell| ^2 + {\cal
   O}(\ell^3)  ,
   \label{trepotenziali}
\end{eqnarray}
with $a_1$, $a_2$, $a_3$ and $b_1$ unknown temperature dependent
coefficients which can be determined using lattice data. In this
paper we want to investigate the general relation between
glueballs and $\ell$ so we will not try to find the best
parameterization to fit the lattice data. In the spirit of the
mean field theory we take $a_2$, $a_3$ and $b_1$ to be positive
constants while $a_1=\alpha (T_{\ast}-T)/T$. With $\ell = |\ell|
e^{i\varphi}$ the extrema are now obtained by differentiating the
potential with respect to $H$, $|\ell|$ and $\varphi$:
\begin{eqnarray}
\frac{\partial V}{\partial H}&=&
\frac{\ln}{2}\left[\frac{eH}{\Lambda^4}\right]+{\cal
P}\left[\ell\right]
=\frac{\ln}{2}\left[\frac{eH}{\Lambda^4}\right]+b_1|\ell|^2=0,  \nonumber  \\
\frac{\partial V}{\partial |\ell|}&=&2|\ell| T^4\left(a_1 +
\frac{H}{T^4} b_1 -3a_3 |\ell|\cos(3\varphi)  +  2a_2 |\ell|^2
\right) =0, \nonumber \\\frac{\partial V}{\partial \varphi}&=&
6|\ell|^3 \, \sin(3\varphi)= 0.  \label{motion3}
\end{eqnarray}
At small temperature the $H/T^4$ term in the second equation
dominates and the solution is $\langle |\ell| \rangle =0$,
$\langle H \rangle=\Lambda^4/e$ and the last equation is verified
for any $\langle \varphi \rangle$, so we choose $\langle \varphi
\rangle = 0$. The second equation can have two more solutions:
\begin{eqnarray}
 \frac{3}{4}\frac{a_3}{a_2} \pm
\sqrt{\frac{9}{16}\frac{a_3^2}{a_2^2} +\frac{\alpha
(T-T_{\ast})}{2 T a_2} - \frac{b_1H}{2a_2 T^4} }  ,
\end{eqnarray}
whenever the square root is well defined (i.e. at sufficiently
high temperatures). The negative sign corresponds to a relative
maximum while the positive one to a relative minimum. We have then
to evaluate the free energy value (i.e. the effective thermal
potential) at the relative minimum and compare it with the one at
$\ell=0$. The temperature value for which the two minima have the
same free energy is defined as the critical temperature and is:
\begin{eqnarray}
T_c = \left[T_{\ast} + \frac{b_1}{e\alpha}\frac{\Lambda^4}{T_c^3}
\right] \frac{\alpha a_2}{\alpha a_2 + a_3^2}  .
\label{criticalTc3}
\end{eqnarray}
{}When $a_3$ vanishes we recover the second order type critical
temperature $T_c$. To derive the previous expression we held fix
the value of $H$ to $\Lambda^4/e$ at the transition point. In a
more refined treatment one should not make such an assumption.
 Below this temperature the minimum is still for
$\langle \ell\rangle=0$ and $\langle H \rangle =\Lambda^4 /e$.

\begin{figure}[h]
\begin{center}
\includegraphics[width=7truecm]{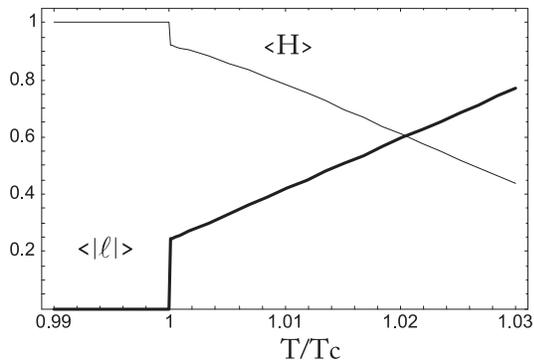}
\end{center}
\caption{A cartoon sketching the gluon condensate $\langle H
\rangle$ normalized to $\Lambda^4/e$ and the $\langle |\ell|
\rangle $ (thick line)  behaviors as function of the temperature.
We have chosen for illustration $a_3=0.3$,  $a_2=1$, $\alpha=2$,
$b_1=0.7$ and $T_c=1.2 \Lambda$.} \label{Figura2}
\end{figure}

Just above the critical temperature the fields jump to the new
values:
\begin{eqnarray}
\langle |\ell|\rangle = \frac{a_3}{a_2} , \qquad \langle H \rangle
=\frac{\Lambda^4}{e} \exp{\left[-2 b_1 \langle |\ell| \rangle^2
\right]}  .
\end{eqnarray}
Close but above $T_c$ (i.e. $T=T_c +\Delta T$) we have:
\begin{eqnarray}
\langle |\ell| \rangle \simeq \frac{a_3}{a_2}+\rho \frac{\Delta
T}{T_c}
 ,
\end{eqnarray}
with \begin{eqnarray} \rho & \simeq& \frac{\alpha a_2}{a_3}
\frac{4 \kappa T_c -3 T_{\ast} }{a_2 T_c -4 b_1 \alpha (\kappa T_c
-
T_{\ast})} , \nonumber \\
\kappa &=& \frac{\alpha a_2 + a_3^2}{\alpha a_2} \ .
\end{eqnarray} a positive function of the coefficients of the
effective potential. In this regime
\begin{eqnarray}
\langle H \rangle = \frac{\Lambda^4}{e} \exp\left[-2
b_1(\frac{a_3}{a_2} + \rho \frac{\Delta T}{T_c})^2 \right]  .
\end{eqnarray}
At high temperature we expect a behavior similar to the one
presented for the two color theory case. A cartoon representing
the behavior of the Polyakov loop and the gluon condensate is
presented in Fig.~\ref{Figura2}.

Since we are in the presence of a first order phase transition
higher order terms in Eq.~(\ref{trepotenziali}) may be important.
However lattice simulations have shown that the behavior of the
Polyakov loop, for 3 colors, resemble a weak first order
transition (i.e. small $a_3$) partially justifying our simple
approach.

The approximation for our coefficients is too crude and it would
certainly be relevant to fit them to lattice simulations.

What we learnt is that the gluon condensate, although not a real
order parameter, encodes the information of the underlying $Z_3$
symmetry. More generally we have shown that once the map between
hadronic states and the true order parameter is known we can use
directly hadronic states to determine when the phase transition
takes place and the order of the phase transition. {}For example
by comparing Fig.~{\ref{Figura1}} and Fig.~\ref{Figura2} we
immediately notice the distinct difference in the gluon condensate
temperature dependence near the phase transition.

\section{Conclusions and Self Criticism}
\label{cinque} Our simple model is able to account for many
features inherent to the Yang-Mills deconfining phase transition.
We related two very distinct and relevant sectors of the theory:
the hadronic sector (the glueballs), and some dimensionless fields
($\ell$) charged under the discrete group $Z_N$ understood as the
center of the underlying $SU(N)$ Yang-Mills theory
\cite{Pisarski:2001pe}.

The gluon-condensate is, strictly speaking, not an order parameter
for the deconfining Yang-Mills phase transition. However we have
shown that the information encoded in the true order parameter
$\ell$ is efficiently communicated to the gluon condensate. Since
the exponential drop of the condensate just above the Yang-Mills
critical temperature is a direct consequence of the behavior of
the true order parameter at the transition we can consider this
drop as a strong signal of deconfinement. This drop has already
been used in various models for the Yang-Mills phase transition.
We have also seen that the reduction in the gluon-condensate is
associated to the increase of the Polyakov loop condensate $\ell$.
The information about the order of the phase transition is also
transferred to the behavior of the gluon condensate. Indeed from
Fig.~\ref{Figura1} and Fig.~\ref{Figura2} we deduce that the drop
is continuous for the gluon condensate in the two color case while
is discontinuous for the three color theory. Physically the
glueballs start decaying into gluons and this information is
encoded in the $H{\cal P}[\ell]$ term of the present theory.

We now better understand the mechanism for transferring
information from the Yang-Mills order parameter to the physical
states.

It is important to stress that our model is very limited since we
neglect the temperature corrections associated with the glueball
gas as well as other dimensionless operators with different
charges under $Z_N$. Finally we did not include any of the excited
glueball and pseudo-scalar glueball states present in the theory.
Besides the temperature dependence of the coefficients in ${\cal
P}[\ell]$ and ${\cal V}[\ell]$ is not known and we have just
adopted the simplest guess consistent with mean-field theory. We
also know that mean-field cannot be the whole story and
corrections need to be included.

It is worth mentioning that the Polyakov loop need not to be the
only acceptable order parameter. For example using an abelian
projection one can define a new  (non local in the cromomagnetic
variables) order parameter \cite{DelDebbio:2002nb}. Our model can
be, in principle, modified to be able to couple the hadronic state
to any reasonable Yang-Mills order parameter.

    The same holds true when considering the introduction of
    quarks. Once identified a true order parameter
    for QCD with quarks we can first construct a model Lagrangian which satisfies the
    ordinary symmetries at zero temperature for the hadronic states
    and then extended it to describe at the same time the order
    parameter and the hadronic states.  Although $\ell$ is not a good order parameter when
    quarks are added to the theory since the $Z_N$ symmetry is explicitly broken
    we can still construct a theory containing $\ell$ and the hadronic states (mesons
    and glueballs) provided we introduce explicit $Z_N$
    breaking terms. This approach might be relevant for
    understanding RHIC physics
    \cite{Scavenius:2001pa,{Scavenius:2002ru}}.

Although the model is at a very early stage of development at the
tree level some of the results are already  fairly robust. For
example the exponential drop of  $\langle H \rangle$ as function
of $\langle \ell \rangle$ is a prediction not expected to be very
sensitive to different sources of corrections. We also note that
the first order behavior of the deconfining three color Yang-Mills
theory is directly inherited by the $\langle H \rangle$. Indeed
this quantity is discontinuous at the phase transition for three
colors while displays a smooth behavior in the two color case. We
expect these results to be quite general. We also stress that they
are connected to the saturation of trace anomaly and the $Z_N$
symmetry at the effective Lagrangian level when considering
simultaneously $\ell$ and $H$.

By computing the temperature dependence of the coefficients in the
present effective theory lattice simulations can test the validity
of the present model while improving the present results.
 \acknowledgments I am indebted to P. Damgaard for enlightening discussions, encouragements and careful reading of
the manuscript. It is a pleasure for me to thank L. Del~Debbio,
J.T.~Lenaghan, M.P. Lombardo, A.D.~Jackson, P. Di~Vecchia,
J.~Schechter and W. Sch\"{a}fer for discussions and reading the
manuscript. This work is supported by the Marie--Curie fellowship
under contract MCFI-2001-00181.

\end{document}